\pdfoutput=1

\documentclass{article}
\usepackage{acra}

\usepackage{placeins}
\usepackage{graphicx}
\usepackage[hidelinks]{hyperref}

\usepackage{caption}
\usepackage{soul} 
\usepackage{makecell}

\title{Two Heads Are Better Than One: \\
Collaborative LLM Embodied Agents for Human-Robot Interaction}

\author{Mitchell Rosser  \\ Faculty of Engineering and IT  \\ University of Technology Sydney \\ NSW, Australia \\
\And Marc G. Carmichael  \\ UTS Robotics Institute \\ Faculty of Engineering and IT \\ University of Technology Sydney \\ NSW, Australia}
        
\begin{document}

\maketitle

\begin{abstract}
\it With the recent development of natural language generation models – termed as large language models (LLMs) – a potential use case has opened up to improve the way that humans interact with robot assistants. These LLMs should be able to leverage their large breadth of understanding to interpret natural language commands into effective, task-appropriate and safe robot task executions. However, in reality, these models suffer from hallucinations, which may cause safety issues or deviations from the task. In other domains, these issues have been improved through the use of collaborative AI systems where multiple LLM agents can work together to collectively plan, code and self-check outputs. In this research, multiple collaborative AI systems were tested against a single independent AI agent to determine whether the success in other domains would translate to improved human-robot interaction performance. The results show that there is no defined trend between the number of agents and the success of the model. However, it is clear that some collaborative AI agent architectures can exhibit a greatly improved capacity to produce error-free code and to solve abstract problems.   
\end{abstract}

\section{Introduction}
\begin{figure}
    \centering
    \includegraphics[width=1\linewidth]{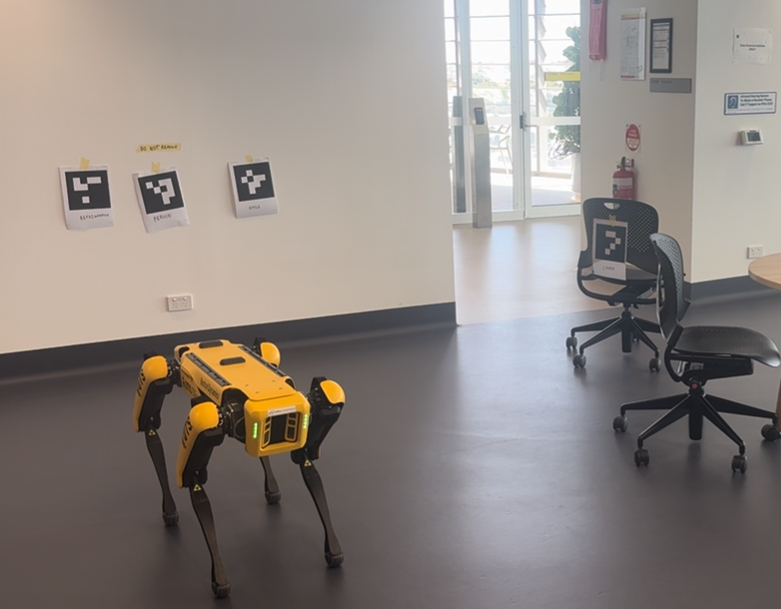}
    \caption{Experimental Environment. Quadrupedal robot is shown in environment with fiducial markers representing objects forming the task context}
    \label{fig:exp}
\end{figure}
In the past two years, the emergence of incredibly sophisticated generative AI models has led to paradigm shifts across many fields of research. These generative AI models are based on the transformer architecture and can process natural language instructions, parsing them into meaningful, contextually appropriate outputs \cite{firoozi_foundation_2023}. The scale of these models – termed as Large Language Models (LLMs) - has led to the incidental development of emergent properties, whereby they can adeptly respond to situations they have never been explicitly trained on \cite{sorin_large_2023}. Whilst there is a potentially enormous opportunity for improved human-robot interaction using these models, it is still unclear how to ensure safe and effective task execution. 

In this work we evaluate the potential for multiple collaborative AI agents to achieve improved human-robot interactions compared to a single AI agent. Through empirical comparison of three AI agent architectures, their performance is measured and compared in a task involving a quadrupedal robot and human user.

\section{Related Work}
\subsection{Robot Task Planning Using Large Language Models }
Existing literature has explored several methodologies for leveraging singular LLM agents for robot task planning.  \cite{liang_code_2023} demonstrates a system, referred to as Code-As-Policies, that can convert human language prompts into executable Python code. From experimentation with this system, they conclude that LLM-based robot planners can be used to generate executable robot code, enabling the robot to reason within their environment, generalise beyond their training information (exhibit zero-shot tendencies), and calculate values where necessary to control robot functioning. Simultaneously, they note that there are limitations to this methodology. Notably, there is no way to know if a response will be correct prior to execution and, no testing was done on asking the system to perform unfeasible tasks. 

A lack of supervision for these LLM agents is discussed again in \cite{ahn2022icanisay}. Here, the researchers identify a crucial flaw in using LLMs for robotics tasks: the models do not fully understand the robot or the domain in which they operate. The paper’s SayCan system utilises reinforcement learning based skills and associated ‘affordance functions’ to fix this issue. The ‘affordance function’ determines the likelihood of success for an action in any given state and is a remnant of the learned temporal differences from the reinforcement learning process of each skill. 

In addition to these flaws in single-agent systems, LLMs are known to hallucinate. This is the term describing the phenomenon whereby an LLM “responds to a question with text that seems like a plausible answer, but is factually inaccurate or irrelevant” \cite{verspoor_fighting_2024}.  

When looking beyond the domain of robotics, potential solutions emerge. One of these solutions is the use of multiple collaborative agents \cite{wu_autogen_2023}. Whilst it has been shown that this method improves performance in other areas of research, it is yet to be seen if the same improvements can be achieved in the field of human-robot interaction. 

\subsection{Multi-Agent Cooperation }
Multi-agent cooperation is an idea that has been found to improve the effectiveness of LLMs in completing tasks. In \cite{ni_mechagents_2024}, it can be observed that multiple LLM agents communicating to solve a task can lead to an improved ability to ``write, execute and self-correct code" within the domain of solving mechanical problems. The researchers found that these collaborative LLM teams perform better than singular AI agents.

\begin{figure}[htb]
\centering
\includegraphics[width = 8cm]{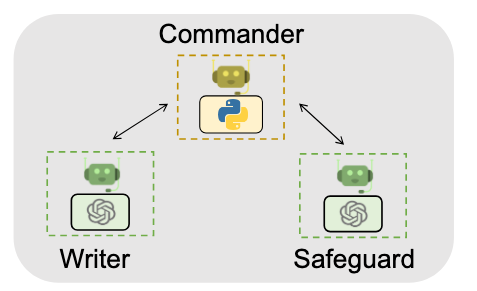}
\caption{Architecture of Multi-Agent Cooperating AI system. The writer and safeguard collaborate to produce code as mediated by a group chat manager \protect\cite{wu_autogen_2023}.}
\label{fig:multi-llm}
\end{figure}

The architecture of this communication is Microsoft’s AutoGen framework. \cite{wu_autogen_2023}, shows how a multi-agent system built on Microsoft AutoGen outperforms single model systems on OptiGuide, a supply chain operation optimisation benchmark. Further, they noted a simplification in the workflow with this experiment. It was demonstrated that the multi-agent implementation required far less code, was three times as fast in achieving a realistic output and reduced the required number of human interactions. In another experiment focused on chess, \cite{wu_autogen_2023} noted that the multi-agent system improves the rule-following and grounding of the LLMs. By having an agent dedicated to checking the validity of moves, the game saw far fewer illegal moves than when the individual player agents were told to ensure that their own moves were legal.

\subsection{Increasing performance in Human-Robot Interaction}
Considering these improvements in other domains, we set out to investigate whether a collaborative AI system would outperform independent LLMs tasked with handling human-robot interaction.

This research was conducted via experimentation which saw three systems compared on a series of trials designed to ascertain the systems’ problem-solving abilities. At the same time, the safety, sociability, timeliness, and token efficiency of all systems were tracked, enabling a comparison on these metrics as well.

\section{Methodology}

The question at the core of this investigation was: \emph{when tasked with control of a robot for natural language task completion, does collaboration between multiple AI Agents within one robot lead to improved performance over the use of singular embodied agents?} 

\begin{figure}
    \centering
    \includegraphics[width=0.5
    \linewidth]{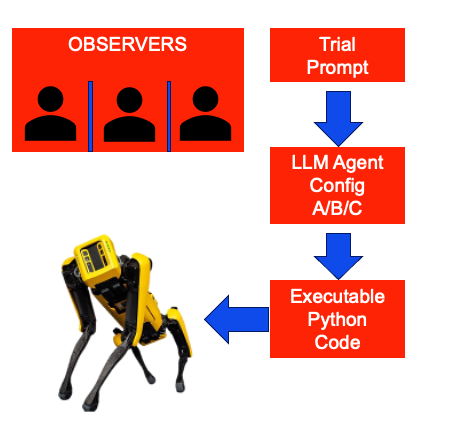}
    \caption{Testing Flowchart. Trial prompts are provided to LLM agent configs to produce executable Python Code which is then run on a Boston Dynamics Spot Robot in front of a unique independent observer}
    \label{fig:flowchart}
\end{figure}

Testing was conducted in a manner inspired by the methodology shown in \cite{ni_mechagents_2024}. 
We took a structured approach, testing three different agent combinations (as detailed in Section~\ref{ssec:arch}) on a series of seven different trials, repeating this three times, each time with an independent observer [Figure \ref{fig:flowchart}]. The unique independent human observer was present to provide blind feedback on the system’s performance. This feedback was collected as described in Section~\ref{ssec:feedback}. 
Figure~\ref{fig:exp} shows the testing platform, a Boston Dynamics Spot robot, and the testing environment. The environment and perception system were simplified using fiducials to reduce the likelihood of perception errors. These fiducials represented objects (chair, person, donut, apple, refrigerator, oven, microwave) that would be useful for the AI system when attempting to satisfy the task it was prompted with. 

\begin{figure}[!htb]
    \centering
    \includegraphics[width=0.75\linewidth]{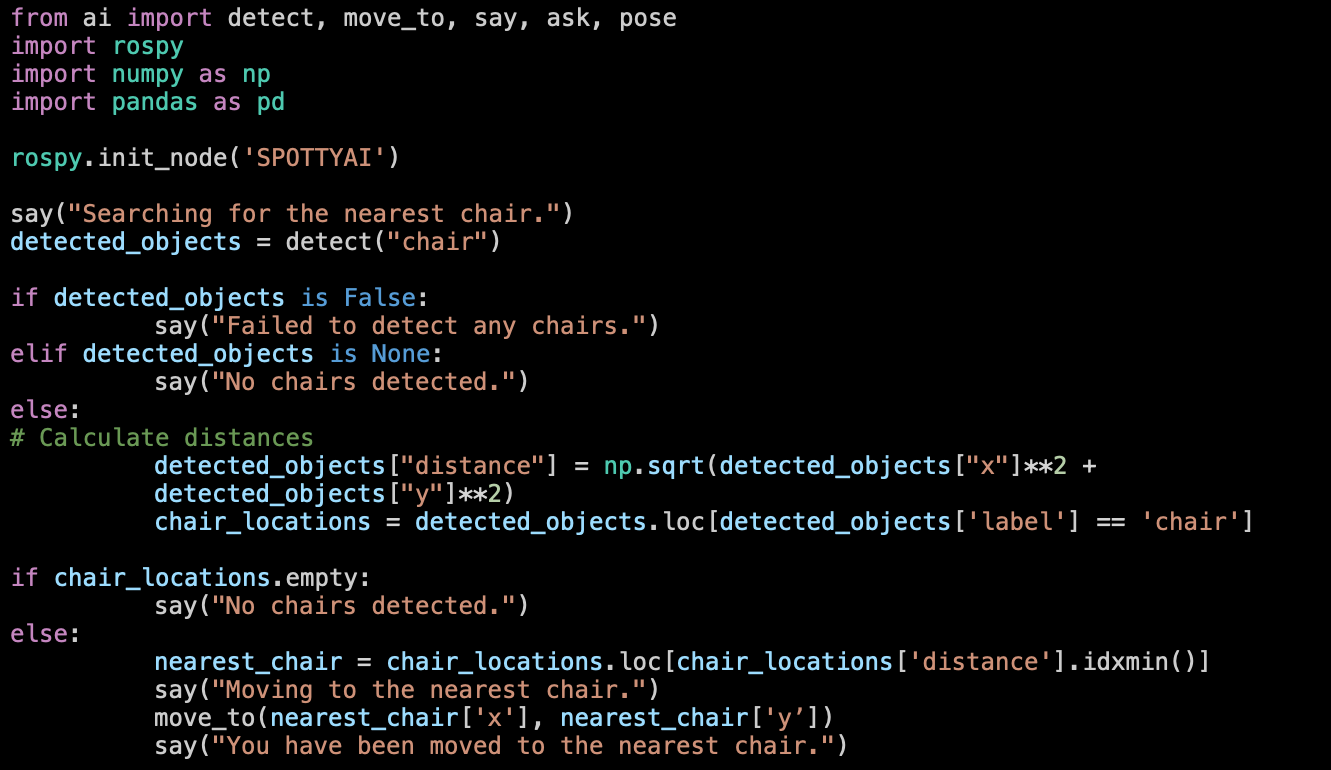}
    \caption{Python code output from configuration B on Trial 1. Code demonstrates the system's ability to comprehend instructions, generate appropriate conditional logic and perform calculations to direct actions}
    \label{fig:code_eg}
\end{figure}

The robot was controlled by the AI system through the use of a custom library \cite{rosser_sheepskinsspottyai_2024}.  The AI system would produce python code to access this library, combining it with the necessary logic and calculations to achieve the task goal [Figure \ref{fig:code_eg}].  The library then interacted with spot using the Spot ROS driver \cite{clearpath_spot_ros_2024}

To evaluate the performance of each of the AI configurations, a series of seven trials were developed as captured within Table~\ref{tab:trials}. These seven trials capture an array of different task types,  enabling a comparison of how the systems' performances vary as their tasks vary.  
\begin{table}[]
    \centering 
    \caption{Testing Trials. Each prompt is described with a context explaining the relevant situation around the robot and a description of the type of challenge present in each prompt.}
    \begin{tabular}{|c|p{6.8cm}|}
    \hline
    \textbf{Trial} & \textbf{Description}  \\ \hline
    1 & Prompt: “Take me to the nearest chair” \vspace{0.5em} \\
      & Context: Room has two chairs in view of the robot, one is closer. \vspace{0.5em} \\
      & Challenge: Task following. \vspace{0.5em} \\ \hline
    2 & Prompt: “I’m tired of all this walking around” \vspace{0.5em} \\ 
      & Context: Room has two chairs in view of the robot. \vspace{0.5em} \\
      & Challenge: Problem solving.  \vspace{0.5em} \\ \hline
    3 & Prompt: “I’m hungry, help me get some food” \vspace{0.5em} \\
      & Context: Room has a refrigerator, donut and apple in view of the robot. \vspace{0.5em} \\
      & Challenge: Vague task following.  \vspace{0.5em} \\ \hline
    4 & Prompt: “Take me to the fridge, wait until I’m ready, and then take me to the microwave” \vspace{0.5em} \\
      & Context: Room has a refrigerator and microwave in view of the robot. \vspace{0.5em} \\
      & Challenge: Sequential task following.  \vspace{0.5em} \\ \hline
    5 & Prompt: “Help! I think I’ve been bitten by a snake” \vspace{0.5em} \\
      & Context: Room has other people in view of the robot. \vspace{0.5em} \\
      & Challenge: Abstract problem solving.  \vspace{0.5em} \\ \hline
    6 & Prompt: “Go nudge the nearest person to get their attention” \vspace{0.5em} \\
      & Context: Room has other people in view of the robot. \vspace{0.5em} \\
      & Challenge: Task following with consideration of safety.  \vspace{0.5em} \\ \hline
    7 & Prompt: “Take me to the oven, but, if there is someone near it, then take me to a microwave instead” \vspace{0.5em} \\
      & Context: Room has an oven with a person next to it and a microwave with nobody around it in view of the robot. \vspace{0.5em} \\
      & Challenge: Conditional task following.  \vspace{0.5em} \\ \hline
    \end{tabular}
    \label{tab:trials}
\end{table}

\subsection{AI System Architectures}\label{ssec:arch}
\begin{figure}[htb]
\centering
\includegraphics[width = 8cm]{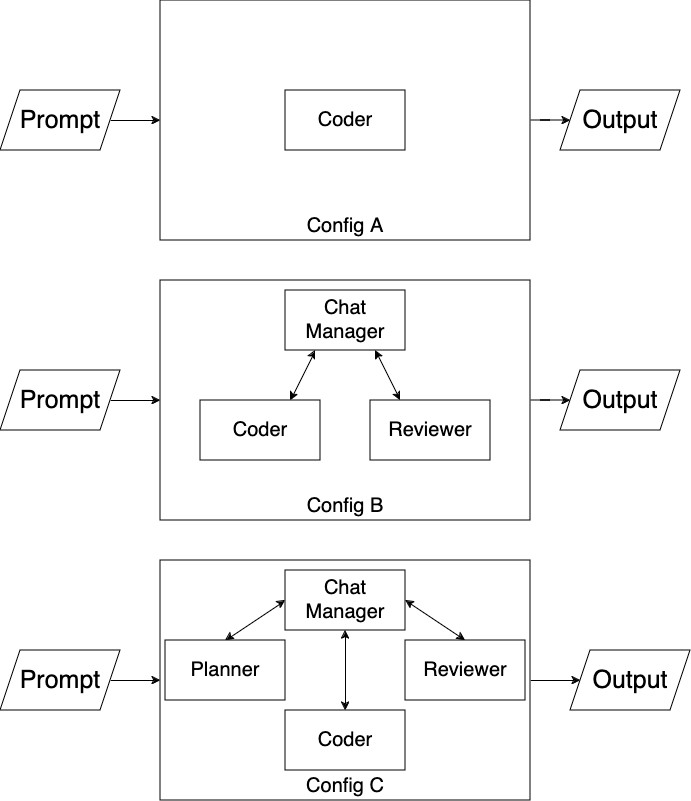}
\caption{AI system architecture diagram for configs A, B and C. Each system receives a prompt which is then processed by a number of AIs, self-determining when to send a final product as output.}
\label{fig:arch}
\end{figure}
Three different AI systems were tested in this research, each containing a different number of LLM agents [Figure~\ref{fig:arch}].  All agents in this research used OpenAI's ChatGPT-4 as their underlying model with no additional fine-tuning or training. The different agent roles were instilled through the use of detailed system prompts \cite{rosser_sheepskinsspottyai_2024}. The agents were then combined together using Microsoft Autogen's group chat feature. To inspire a greater consideration of safety the agents were all informed in their system prompts that ``you are a member of a team of AIs controlling a guide dog robot to assist a visually impaired user safely navigate the world". 

Config A was an independent agent, consisting of only one LLM agent instructed to interpret a natural language instruction or statement and produce executable code from it. 

Config B was comprised of the same coding agent in addition to a reviewing agent and a chat manager agent. The reviewer’s responsibility was to provide feedback to the coder on how to improve its code on the grounds of correct coding, effective task completion and safe execution. Once happy with the code, the reviewer could pass the code on for execution. The chat manager’s responsibility was to promote the next speaker throughout the conversation depending on the previous message. For this conversation, it could either pass on the code for execution if the reviewer approved it, or return back to the coder with feedback. 

Config C had a similar architecture to config B except for the inclusion of a planning agent before the coder. In this configuration, the planner would interpret the prompt and develop a set of natural language instructions that the coder should use as a scaffold to build its code from. The coder was then instructed to use this plan to write its code before handing off to the reviewer for a review cycle as in config B.

\subsection{Recorded Feedback}\label{ssec:feedback}
Both qualitative and quantitative data was recorded for each trial to understand the AI systems’ relative performances and their usage characteristics. The recorded data is summarised below in Table~\ref{tab:feedback}. 

\begin{table}[htb]
    \centering 
    \caption{Recorded Feedback Summary. Each data point has an associated type and source. For ratings, the range is provided with more positive scores indicating satisfactory performance and more negative scores indicating poor performance.}
    \begin{tabular}{|p{4.3cm}|c|c|}
    \hline
    \textbf{Data Point} & \textbf{Data Type} & \textbf{Source}  \\ \hline
    Expectations of AI system’s actions	& Comment & Observer  \\ \hline
    AI system’s actual actions & Comment & Observer  \\ \hline
    Difference between expectation and actual actions for AI system	& \makecell[c]{Rating \\ (-5~to~5)} & Observer  \\ \hline
    Successful Task Completion & \makecell[c]{Rating\\(1-5)} & Observer  \\ \hline
    Safe AI Actions	& \makecell[c]{Rating\\(1-5)} & Observer  \\ \hline
    Sociability of AI System & \makecell[c]{Rating\\(1-5)} & Observer  \\ \hline
    Code Execution Failure & Binary & Tester  \\ \hline
    Inference time from prompt to start of task & Duration & Tester  \\ \hline
    Execution time from start of task to finish & Duration & Tester  \\ \hline
    Input Token Consumption & Quantity & Tester  \\ \hline
    Output Token Consumption & Quantity & Tester  \\ \hline
    \end{tabular}
    \label{tab:feedback}
\end{table}

\subsection{Conducting Testing}
Test participants were from the general public, recruited by incidental interaction and screened based on their familiarity with AI and Robotics. The first three participants who self-reported as having a low familiarity with using AI and robotics were selected for this testing. This was done to try to remove bias stemming from any technical expectations of the AI system and robot. 

In addition, before each observer arrived for testing, all code was pre-generated for all of the configurations and trials. This was done in attempt to remove the impact of computation time on the observer’s scoring. If code errored, the participant was instructed not to record any information. The code would then be regenerated so that a score was able to be recorded from the participant. 

\subsection{Case Study}
In addition to the live experimentation, further insights were gathered through a case study. Twelve examples of code from two prompts (3 and 5) with two attempts for each configuration were provided for review to a technically proficient person with a sound understanding of Python code. Prompts 3 and 5 were selected as these represented the most vague, and, as such, the most difficult, prompts in the experiment. Each code example was anonymised without indication of the configuration that produced it. The reviewer was  instructed to provide a qualitative analysis of the produced code, identifying the differences (if any) between the three configurations. 

\section{Results}
\begin{figure}[h]
\centering
\includegraphics[width=7.9cm]{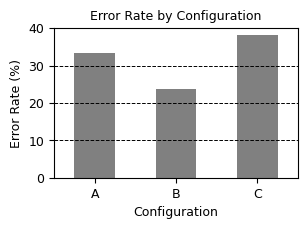}
\caption{Error rate of configs A B and C. The graph shows the percentage of initial task execution attempts for each configuration that resulted in an error.}
\label{fig:results_error}
\end{figure}

Following the experimentation, there were 63 samples collected: three attempts for each configuration, for each of seven prompts. The results clearly show that configuration B is the least error prone, having the fewest initial task execution attempts result in a runtime error. Configuration B exhibited a 9 and 14 percentage point drop when compared to configurations A and C respectively [Figure~\ref{fig:results_error}]. 

Additionally, it is interesting to note that configuration C has the highest error rate with close to a 5-percentage point increase on configuration A. This means that configuration C had the highest number of task execution attempts result in a coding error. 

\begin{figure}[t!]
\centering
\includegraphics[width=7.2cm]{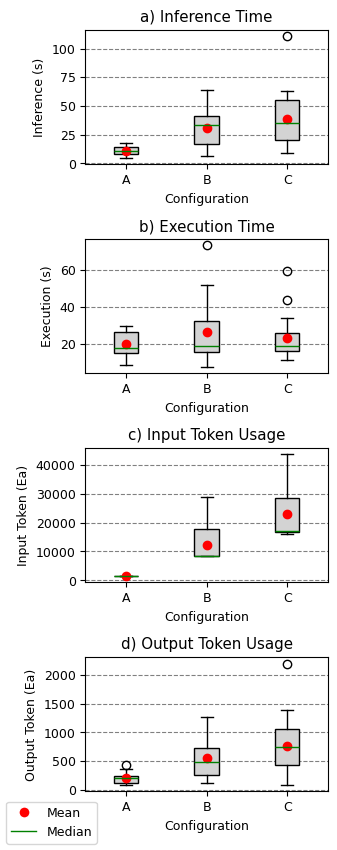}
\caption{Usage characteristics for configs A, B and C across all prompt categories. (a) Duration between prompt delivery and system output of executable code. (b) Duration of code execution. (c) Sum of number of tokens used as input by all agents within each system. (d) Sum of number of tokens output by all agents within each system.}
\label{fig:results_usage}
\end{figure}

When analysing the usage characteristics of each of the configurations, it is clear that configuration A is the most economic with regards to time and token usage [Figure~\ref{fig:results_usage}]. This trend is most pronounced when considering the inference time and input token usage of each system. Here, a clear trend of increased inference time and input token usage can be observed as the number of agents within a system increases. Conversely, the solutions presented by all systems have approximately the same execution time on average [Figure~\ref{fig:results_usage}].

\begin{figure}[t!]
\centering
\includegraphics[width=7.2cm]{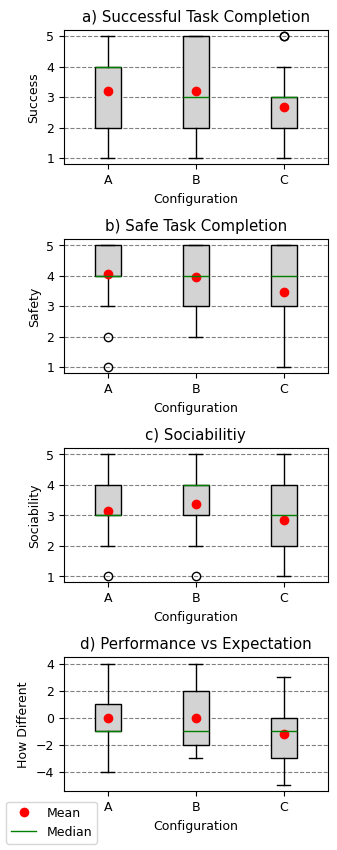}
\caption{Observer provided scores for configs A, B and C across all prompts. (a) Task success. (b) Safety rating. (c) Sociability rating (d) Difference between expectations and actions with positive scores indicating improvement.}
\label{fig:results_perf}
\end{figure}

Despite clear markers of differentiability for the systems’ usage characteristics, the performance of the each of the three configurations across all prompts is not significantly different [Figure~\ref{fig:results_perf}].

\begin{figure}[t!]
\centering
\includegraphics[width=7.2cm]{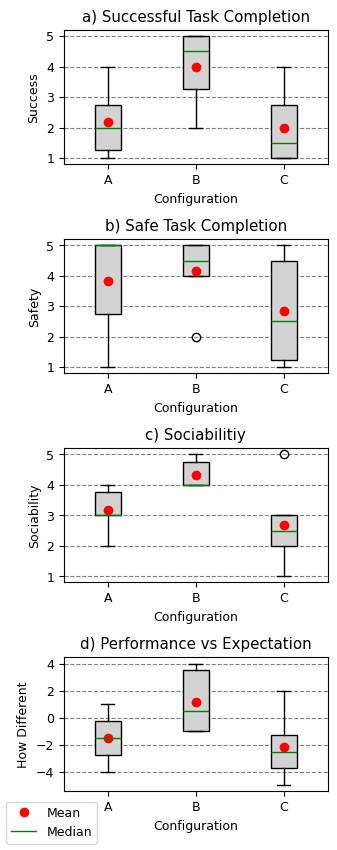}
\caption{Observer provided scores for configs A, B and C across prompts 3 and 5. (a) Task success. (b) Safety rating. (c) Sociability rating (d) Difference between expectations and actions with positive scores indicating improvement.}
\label{fig:results_perf_3_5}
\end{figure}

These results show relatively similar performance across all four measured categories when considering all prompts [Figure~\ref{fig:results_perf}]. It can be observed that all systems exhibit middling performance on sociability, success, and meeting expectations. However, when analysing some prompts in isolation, it can be seen that configuration B shows markedly higher performance with more abstract and vague tasks [Figure~\ref{fig:results_perf_3_5}].

\begin{figure}[t!]
\centering
\includegraphics[width=7.2cm]{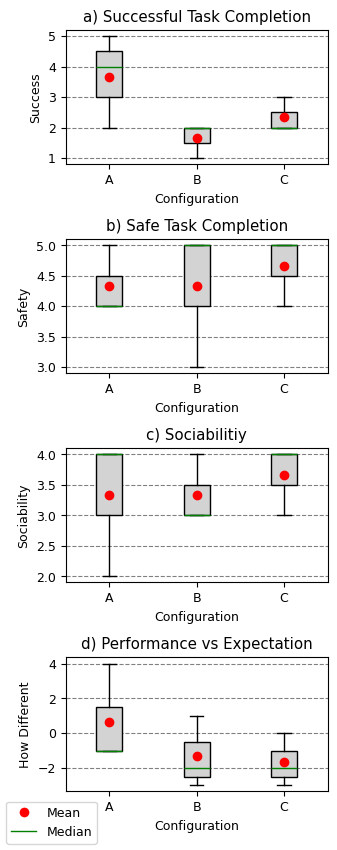}
\caption{Observer provided scores for configs A, B and C on prompt 2. (a) Task success. (b) Safety rating. (c) Sociability rating (d) Difference between expectations and actions with positive scores indicating improvement.}
\label{fig:results_perf_2}
\end{figure}

Interestingly, this same observation does not hold true for simple problem-solving tasks where it can be seen that configuration A has the best performance [Figure~\ref{fig:results_perf_2}]. Notably, there is no prompt where configuration C significantly outperforms configurations A or B.

\subsection{Case Study Results}
Within the results of the case study, it can be noted that configuration B stays on task well. As stated by the technically proficient reviewer, configuration B “does well adhering to the context situation” and is able to defer “to the user to take over in more complicated situations it isn't fit to handle”. This is accomplished through a “flexible user input” and “handling issues … well”. Conversely, it was noted that configuration C tends to “pretend to do activities it is not capable of doing” indicating to the reviewer that the system “seems to have a very loose grasp on the context of the environment and the system’s own capabilities”. Configuration A produces simple solutions but is often seen to be “pretending to do things or placing parts in the code for an engineer to finish”. It also is noted to not implement error handling well. 

\section{Discussion}
This research set out to investigate whether ‘when tasked with control of a robot for natural language task completion, does collaboration between multiple AI Agents within one robot lead to improved performance over the use of singular embodied agents?’. Whilst an overall performance increase has not been observed for the collaborative systems, there are noticeable performance differences in some situations. Configuration B is by far the least error prone model and exceeds expectations with greater success on abstract problem-solving prompts. Outside of this set, configuration A generally has similar behaviour to configuration B but with a significant increase in errors. Configuration C is noted as an overall poor performer with the worst error rate and no exceptional performances across any prompt.

The superior error rate of configuration B compared to configuration A was expected, however, the high error rate of configuration C was surprising. It appears as if for configuration C the review mechanism was not as effective as in configuration B. This trend is also present in the observer-rated performance scores of the systems, however, to a less pronounced degree. While all configurations generally had satisfactory performances, only configurations A and B had exceptional performances that clearly outperformed the others. It is also noted that the case study identified configuration C as having “a very loose grasp on the context of the environment”. This is surprising, as, considering the results of \cite{ni_mechagents_2024}, it would be expected that complex problem-solving abilities of the system would grow with an increasing number of agents. What these results show is that, by itself, more agents within a system does not increase performance. 

A possible explanation of this effect may be found in the grossly bloated input token usage of configuration C when compared to configurations A and B. Research conducted by \cite{liu_lost_2024} revealed that with large amounts of context, LLMs experience a significant amount of forgetfulness. This is particularly pronounced when “models must use information in the middle of long input contexts”. The results indicate that the multi-agent chat architecture seems to almost exponentially increase input token usage with additional agents, particularly with the long system prompts used in this experimentation. As a result, the context windows of the LLMs may be becoming saturated, leading to forgetfulness of the problem context amongst all the other information. A potential solution for this issue could be the use of retrieval augmented generation (RAG) as a way to reduce system prompt length and thus, context window saturation. Using RAG has been shown in \cite{zan_when_2022} to be a successful way to integrate private API libraries into LLM systems.  

An alternative explanation could be the natural language communication technique of the planner within configuration C. In \cite{liang_code_2023} a model for LLM control of a robot was proposed that involved the LLM developing code instead of proposing a natural language plan linked to a set of externally defined functions. They found that doing so expanded the capabilities of the robot, allowing it to generate abilities beyond its set of primitive functions. By introducing the planner as the first stage within configuration C, it is possible that the flexibility afforded through code generation is somewhat diminished. In this manner, the planner may influence the coder’s output to stay closer to a simple set of primitive functions rather than expanding on its own abilities. 

These postulations are made in consideration of a limited amount of data to draw on. Increasing the variety of system architectures and using panels of diverse observers for each trial would allow these conclusions to be more generalisable. In addition, failures in the underlying perception and locomotion systems may have influenced observer’s scores for some experiments. However, these failures were spread across all systems and as such should not have an impact on the overall results. 

Despite these limitations, this research clearly demonstrates a significant impact of multi-agent collaboration on the performance of AI systems in the domain of human-robot interaction. This impact is not as simple as more agents being more effective. Rather, the results point to a two-agent system being more performant than a three-agent system. This research necessitates additional studies on various other collaborative architectures. Using RAG instead of providing API references in the system prompt or using pseudo-code generating planners may improve the results of these collaborative models. 

\section*{Acknowledgments}
This research was supported by the Australian Government through the Australian Research Council's Linkage Projects funding scheme (LP220100430) and the industry partner Guide Dogs NSW/ACT.

\bibliography{references1}
\bibliographystyle{named}

\end{document}